# Radiation protection at CERN

*Doris Forkel-Wirth, Stefan Roesler, Marco Silari, Marilena Streit-Bianchi, Christian Theis, Heinz Vincke, and Helmut Vincke*
CERN, Geneva, Switzerland

**Abstract**
This paper gives a brief overview of the general principles of radiation protection legislation; explains radiological quantities and units, including some basic facts about radioactivity and the biological effects of radiation; and gives an overview of the classification of radiological areas at CERN, radiation fields at high-energy accelerators, and the radiation monitoring system used at CERN. A short section addresses the ALARA approach used at CERN.

## 1  Introduction

CERN's radiation protection policy stipulates that the exposure of persons to radiation and the radiological impact on the environment should be as low as reasonably achievable (the ALARA principle), and should comply with the regulations in force in the Host States and with the recommendations of competent international bodies. This paper gives a brief overview of the general principles of radiation protection legislation; explains radiological quantities and units, including some basic facts about radioactivity and the biological effects of radiation; and gives an overview of the classification of radiological areas at CERN, radiation fields at high-energy accelerators, and the radiation monitoring system used at CERN. Finally, a short section addresses the ALARA approach used at CERN.

## 2  General principles of radiation protection legislation

The International Commission on Radiological Protection (ICRP) has specified in its Recommendation 60 [1] that any exposure of persons to ionizing radiation should be controlled and should be based on three main principles, namely:

– *justification:* any exposure of persons to ionizing radiation has to be justified;

– *limitation:* personal doses have to be kept below legal limits;

– *optimization:* personal and collective doses have to be kept as low as reasonably achievable (ALARA).

These recommendations have been fully incorporated into CERN's radiation safety code [2].

## 3  Radiological quantities and units [3]

It would be desirable if the legal protection limits could be expressed in directly measurable physical quantities. However, this does not allow the biological effects of exposure of the human body to ionizing radiation to be quantified. For this reason, protection limits are expressed in terms of so-called protection quantities, which, although calculable, are not measurable. Protection quantities quantify the extent of exposure of the human body to ionizing radiation from both whole-body and partial-body external irradiation and from the intake of radionuclides. In order to demonstrate compliance with dose limits, so-called operational quantities are typically used, which are aimed at providing conservative estimates of protection quantities. The radiation protection detectors used for individual and area monitoring are often calibrated in terms of operational quantities.

## 3.1 Physical quantities

The *fluence* $\Phi$ (measured in units of 1/m$^2$) is the quotient of d$N$ by d$a$, where d$N$ is the number of particles incident upon a small sphere of cross-sectional area d$a$:

$$\Phi = \frac{dN}{da}. \qquad (1)$$

In dosimetric calculations, the fluence is frequently expressed in terms of the lengths $l$ of particle trajectories. It can be shown that the fluence is also given by

$$\Phi = \frac{dl}{dV}, \qquad (2)$$

where d$l$ is the sum of the particle trajectory lengths in the volume d$V$.

The *absorbed dose D* (measured in units of grays; 1 Gy = 1 J/kg = 100 rad) is the energy imparted by ionizing radiation to a volume element of a specified material divided by the mass of that volume element.

The *kerma K* (measured in units of grays) is the sum of the initial kinetic energies of all charged particles liberated by indirectly ionizing radiation in a volume element of a specified material divided by the mass of that volume element.

The *linear energy transfer L* or LET (measured in units of J/m, but often given in keV/μm) is the mean energy d$E$ lost by a charged particle owing to collisions with electrons in traversing a distance d$l$ in matter. Low-LET radiation ($L < 10$ keV/μm) comprises X-rays and gamma rays (accompanied by charged particles due to interactions with the surrounding medium), and light charged particles such as electrons that produce sparse ionizing events far apart on a molecular scale. High-LET radiation ($L > 10$ keV/μm) comprises neutrons and heavy charged particles that produce ionizing events densely spaced on a molecular scale.

The *activity A* (measured in units of becquerels; 1 Bq = 1/s = 27 pCi) is the expectation value of the number of nuclear decays in a given quantity of material per unit time.

## 3.2 Protection quantities

The *organ absorbed dose $D_T$* (measured in units of grays) in an organ or tissue $T$ of mass $m_T$ is defined by

$$D_T = \frac{1}{m_T} \int_{m_T} D \, dm. \qquad (3)$$

The *equivalent dose $H_T$* (measured in units of sieverts; 1 Sv = 100 rem) in an organ or tissue $T$ is equal to the sum of the absorbed doses $D_{T,R}$ in an organ or tissue caused by different radiation types $R$ weighted by so-called radiation weighting factors $w_R$:

$$H_T = \sum_R w_R * D_{T,R}. \qquad (4)$$

This expresses the long-term risks (primarily cancer and leukaemia) from low-level chronic exposure. The values of $w_R$ recommended by the ICRP [4] are unity for photons, electrons, and muons, 2.0 for protons and charged pions, 20.0 for ions, and a function of energy for neutrons (of energy $E_n$):

$$w_R = \begin{cases} 2.5 + 18.2 * e^{-\left[\frac{\ln(E_n)^2}{6}\right]} & \text{if } E_n < 1 \text{ MeV}, \\ 5.0 + 17.0 * e^{-\left[\frac{\ln(2*E_n)^2}{6}\right]} & \text{if } 1 \text{ MeV} < E_n < 50 \text{ MeV}, \\ 2.5 + 3.25 * e^{-\left[\frac{\ln(0.04*E_n)^2}{6}\right]} & \text{if } E_n < 50 \text{ MeV}. \end{cases} \quad (5)$$

The *effective dose E* (measured in units of sieverts) is the sum of the equivalent doses, weighted by the tissue weighting factors $w_T$ (where $\sum_T w_T = 1$), for several organs and tissues $T$ of the body that are considered to be the most sensitive [4]:

$$E = \sum_T W_T * H_T. \quad (6)$$

### 3.3 Operational quantities

The *ambient dose equivalent H\*(10)* (measured in units of sieverts) is the dose equivalent at a point in a radiation field that would be produced by a corresponding expanded and aligned field in a 30 cm diameter sphere of tissue of unit density at a depth of 10 mm, on the radius vector opposite to the direction of the aligned field. The ambient dose equivalent is the operational quantity for area monitoring.

The *personal dose equivalent $H_p(d)$* (measured in units of sieverts) is the dose equivalent in standard tissue at an appropriate depth $d$ below a specified point on the human body. The specified point is normally taken to be where an individual dosimeter is worn. The personal dose equivalent $H_p(10)$, with a depth $d = 10$ mm, is used for the assessment of the effective dose, and $H_p(0.07)$, with $d = 0.07$ mm, is used for the assessment of doses to the skin and to the hands and feet. The personal dose equivalent is the operational quantity for monitoring of individuals.

### 3.4 Dose conversion coefficients

Dose conversion coefficients allow the direct calculation of protection or operational quantities from the particle fluence and are functions of the particle type, energy, and irradiation configuration. The most commonly used coefficients are those for the effective dose and ambient dose equivalent. The former are based on simulations in which the dose to organs of anthropomorphic phantoms is calculated for approximate actual conditions of exposure, such as irradiation of the front of the body (antero-posterior irradiation) or isotropic irradiation. Dose conversion coefficients from fluence to effective dose for antero-posterior irradiation are shown in Fig. 1.

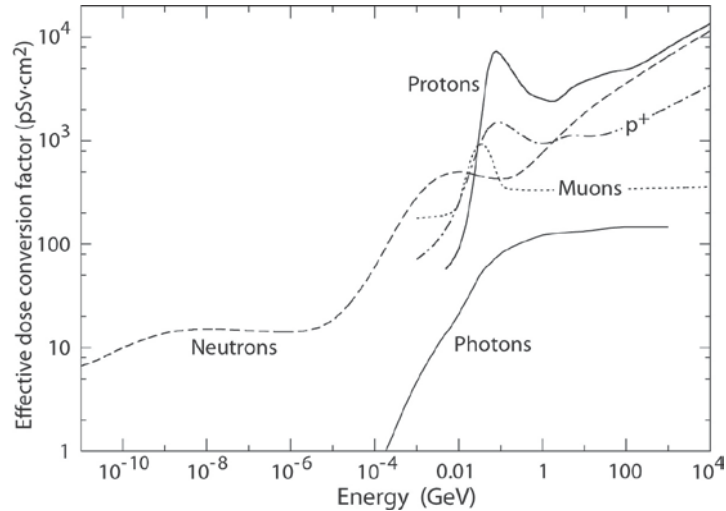

**Fig. 1:** Conversion coefficients from fluence to effective dose for antero-posterior irradiation

# 4 Health effects of ionizing radiation

Radiation can cause two types of health effects, deterministic and stochastic.

*Deterministic effects* are tissue reactions which cause injury to a population of cells if a given threshold of absorbed dose is exceeded. The severity of the reaction increases with dose. The quantity used for tissue reactions is the absorbed dose $D$. When particles other than photons and electrons (low-LET radiation) are involved, a dose weighted by the Relative Biological Effectiveness (RBE) may be used. The RBE of a given radiation is the reciprocal of the ratio of the absorbed dose of that radiation to the absorbed dose of a reference radiation (usually X-rays) required to produce the same degree of biological effect. It is a complex quantity that depends on many factors such as cell type, dose rate, and fractionation.

*Stochastic effects* are malignant diseases and inheritable effects for which the probability of an effect occurring, but not its severity, is a function of dose without a threshold.

## 4.1 Biological effects

The biological effect of radiation depends on the type and energy of the radiation (photons, neutrons, protons, heavy nuclei, etc.), on whether the irradiation is external or internal, on whether it is from radionuclides inhaled or ingested, and on the dose and dose rate received. Furthermore, the type of organ irradiated (for example, the bone marrow is much more sensitive than the liver) and whether local or total body irradiation has occurred will strongly affect the severity and outcome of the damage produced. All this explains the need for and use of various weighting factors to derive equivalent and effective doses in radiation protection.

The cascade of reactions and interactions that occurs when radiation hits a biological system is a mixture of direct and indirect effects, each of them occurring on a different time-scale. The damage starts with the direct ionization and excitation of biological molecules or the creation of free radicals, which gives rise to peroxides, and the interaction of these with DNA molecules produces both repairable and non-repairable damage. Breaks in DNA single strands are highly repairable, but the problem is to know how much misrepair will occur for various doses and types of radiation. In fact, misrepair can either induce programmed cell death, called apoptosis, or produce non-lethal mutations. The damage will result either in deterministic effects (cell death, necrosis, or damage to tissues, organs, or the body, etc.) or in stochastic effects. The latter, resulting from non-lethal mutations, may become visible only many years after irradiation as a cancer or, if the germ cells have been affected, it may be transmitted to future generations in the form of inheritable damage.

The dose for which 50% of individuals will die within 30 days after acute irradiation exposure ($LD_{50/30}$) is 2.5 to 4.5 Gy. More recently, the doses for which 10% and 90% of the population may die from acute irradiation have been estimated; these values are 1–2 Gy for $LD_{10}$ and ~5–7 Gy for $LD_{90}$, respectively.

For each type of deterministic effect (erythraemia, depletion of bone marrow and blood cells, necrosis, vomiting, etc.), there is a dose threshold for the damage to become assessable or visible. The various types of damage observable after acute irradiation, and their dose equivalents are listed in Table 1.

In spite of the long controversy about the presence or absence of damage at extremely low doses less than 0.2 Gy, the absence of a threshold for the stochastic effects is generally accepted. Based on such an assumption, the probability of risk at extremely low doses has been calculated and applied to set occupational and public dose limits for radiation protection. More detailed information about the biological effects of ionizing radiation is given in Ref. [6].

**Table 1:** Radiation damage to the human body [5]

| Dose (whole-body irradiation) | Effects |
| --- | --- |
| <0.25 Gy | No clinically recognizable damage |
| 0.25 Gy | Decrease in white blood cells |
| 0.5 Gy | Increasing destruction of leukocyte-forming organs (causing decreased resistance to infections) |
| 1 Gy | Marked changes in the blood (decrease in the numbers of leukocytes and neutrophils) |
| 2 Gy | Nausea and other symptoms |
| 5 Gy | Damage to the gastrointestinal tract causing bleeding and ~50% death |
| 10 Gy | Destruction of the neurological system and ~100% death within 24 h |

## 5   Radiation levels [3]

– *Natural background radiation*. On average, worldwide, the annual whole-body dose equivalent due to all sources of natural background radiation ranges from 1.0 to 13 mSv, with an average of 2.4 mSv [7]. In certain areas, values up to 50 mSv have been measured. A large fraction (typically more than 50%) originates from inhaled natural radioactivity, mostly radon and radon decay products. The dose equivalent due to radon can vary by more than one order of magnitude: it is 0.1–0.2 mSv per year in open areas, 2 mSv per year on average in houses, and more than 20 mSv per year in poorly ventilated mines.

– *Cosmic ray background radiation*. At sea level, the whole-body dose equivalent due to cosmic ray background radiation is dominated by muons; at higher altitudes, nucleons also contribute. The dose equivalent rates range from less than 0.1 μSv/h at sea level to a few μSv/h at aircraft altitudes.

– *Cancer induction*. The cancer induction probability is about 5% per sievert on average for the entire population [4].

– *Lethal dose*. The whole-body dose from penetrating ionizing radiation resulting in 50% mortality in 30 days, assuming no medical treatment, is 2.5–4.5 Gy (RBE-weighted when necessary), as measured internally on the longitudinal centre line of the body. The surface dose varies because of variable body attenuation and may be a strong function of energy.

– *Recommended dose limits*. The ICRP recommends a limit for radiation workers of 20 mSv effective dose per year averaged over five years, with the provision that the dose should not exceed 50 mSv in any single year [4]. The limit in the EU countries and Switzerland is 20 mSv per year; in the US, it is 50 mSv per year (or 5 rem per year). Many physics laboratories in the US and elsewhere set lower limits. The dose limit for the general public is typically 1 mSv per year.

## 5.1 Radiation levels in Switzerland

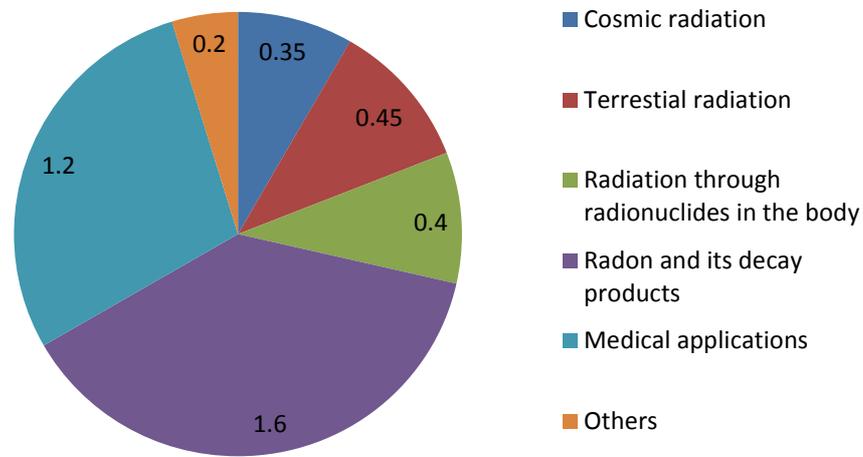

**Fig. 2:** Mean radiation exposure in Switzerland per year (in mSv) [8]

The contributions to the mean radiation exposure in Switzerland [8] are given in Fig. 2. However, the contribution of radon to the total radiation exposure varies strongly in Switzerland. This is determined mainly by the amount of natural radon (which is a product of the natural decay of uranium and thorium) in the soil. A radon map of Switzerland is shown in Fig. 3.

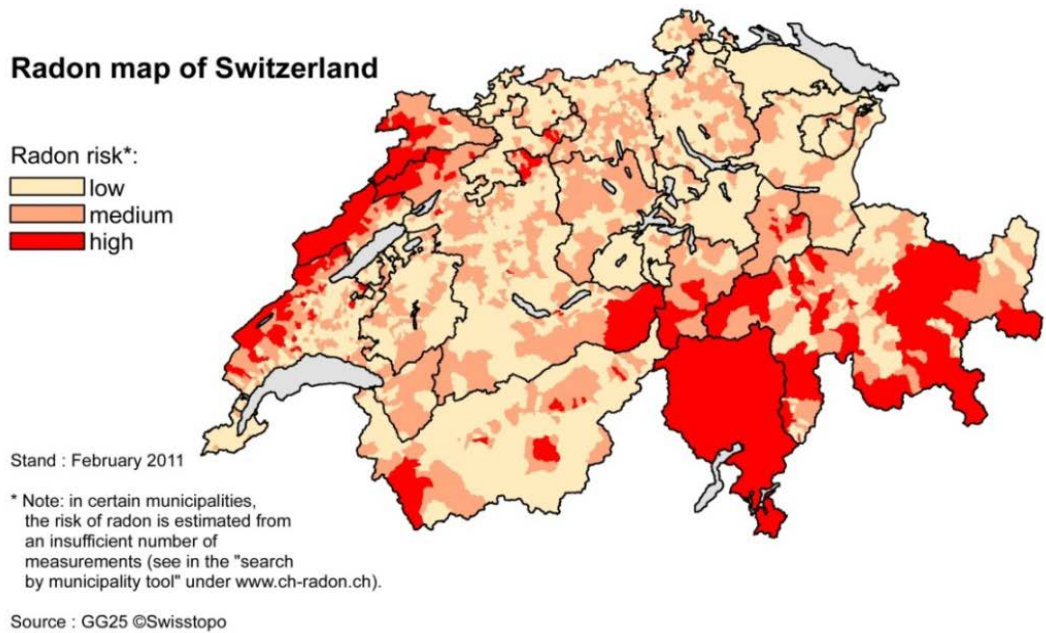

**Fig. 3:** Radon risk in Switzerland. The map is based on measurements performed in buildings (occupied rooms) [9].

## 6  Radiological classification of CERN's areas, and dose limits

### 6.1  Radiological classification at CERN

The areas inside CERN's perimeter are classified as a function of the effective dose a person is liable to receive during his stay in the area under normal working conditions during routine operation. In line with Safety Code F (2006) [2], three types of areas are distinguished:

– Non-designated Areas;
– Supervised Radiation Areas;
– Controlled Radiation Areas.

The latter two are jointly termed Radiation Areas.

The potential external and internal exposures have to be taken into account when assessing the effective dose that persons may receive when working in an area under consideration. Limitation of exposure in terms of effective dose is ensured by limiting an operational quantity, the ambient dose equivalent rate $H^*(10)$ for exposure from external radiation, and by setting action levels for airborne radioactivity and specific surface contamination at the workplace for exposure from incorporated radionuclides. The radiological classification used at CERN is shown in Table 2.

**Table 2:** Synopsis of the classification of Non-designated Areas and Radiation Areas at CERN

| Area | Dose limit [year] | Ambient dose equivalent rate — Work place | Ambient dose equivalent rate — Low occupancy | Sign |
|---|---|---|---|---|
| Non-designated | 1 mSv | 0.5 µSv/h | 2.5 µSv/h | |
| Supervised | 6 mSv | 3 µSv/h | 15 µSv/h | Dosimeter obligatory |
| Simple | 20 mSv | 10 µSv/h | 50 µSv/h | Dosimeter obligatory |
| Limited Stay | 20 mSv | | 2 mSv/h | LIMITED STAY / SÉJOUR LIMITÉ — Dosimeters obligatory |
| High Radiation | 20 mSv | | 100 mSv/h | HIGH RADIATION / HAUTE RADIATION — Dosimeters obligatory |
| Prohibited | 20 mSv | | > 100 mSv/h | PROHIBITED AREA / ZONE INTERDITE — No Entry |

(Supervised is a Radiation Area. Simple, Limited Stay, High Radiation, and Prohibited are Controlled Areas within the Radiation Area.)

### 6.2  Dose limits and classification of workers

All occupationally exposed persons at CERN are classified into one of two categories:

*Category A:* persons who may be exposed in the exercise of their profession to more than 3/10 of the limit in terms of effective dose in 12 consecutive months;

*Category B:* persons who may be exposed in the exercise of their profession to less than 3/10 of the limit in terms of effective dose in 12 consecutive months.

The CERN dose limits are compliant with those of most European countries or even more restrictive. Examples of the dose limits in some European countries are given in Table 3.

Table 3: Dose limits at CERN and in some European countries

|  | Dose limits for 12 consecutive months (mSv) | | |
| --- | --- | --- | --- |
|  | Non-occupationally exposed persons | Occupationally exposed persons | |
|  |  | Category B | Category A |
| EURATOM members | 1 | 6 | 20 |
| Germany and France | 1 | 6 | 20 |
| CERN | 1 | 6 | 20 |
| Switzerland | 1 | 20 | |

### 6.3 CERN's limits for radionuclides of artificial and natural origin

At CERN, material is considered as radioactive if one or more of the following three criteria are fulfilled.

#### 6.3.1 Specific activity and total activity

CERN's Safety Code F [2] applies to any practice involving material containing radionuclides for which

- the specific activity exceeds the CERN exemption limits [10]; and
- the total activity exceeds the CERN exemption limits [10].

For material containing a mixture of radionuclides of artificial origin, the following sum rule is applied to exempt it from any further regulatory control:

$$\sum_{i=1}^{n} \frac{a_i}{LE_i} < 1, \qquad (7)$$

where $a_i$ is the specific activity (Bq/kg) or the total activity (Bq) of the $i$-th radionuclide of artificial origin in the material, $LE_i$ is the CERN exemption limit for that radionuclide, and $n$ is the number of radionuclides present.

#### 6.3.2 Dose rate

CERN's Safety Code F [2] applies to all materials for which the ambient dose equivalent rate measured at a distance of 10 cm from the item exceeds 0.1 µSv/h after subtraction of the background.

#### 6.3.3 Surface contamination

CERN's Safety Code F [2] applies to all materials for which the surface contamination exceeds 1 Bq/cm$^2$ in the case of unidentified beta and gamma emitters and 0.1 Bq/cm$^2$ in the case of unidentified alpha emitters. Once a radionuclide has been identified, specific CERN CS-values [10] can be used, and the following sum rule should be applied:

$$\sum_{i=1}^{n} \frac{c_i}{CS_i} < 1, \qquad (8)$$

where $c_i$ is the value of the surface contamination (Bq/cm$^2$) of the $i$-th radionuclide, $CS_i$ is its CS-value, and $n$ is the number of identified radionuclides.

## 7 Induced radioactivity [11]

Neutrons are not affected by the Coulomb barrier of nuclei, and can thus react at any energy and produce radioactive nuclides. Neutron capture dominates for thermal neutrons, whereas reactions of type (n, p), (n, α), (n, 2n), etc. occur with increasing energy. High-energy neutrons cause spallation reactions that can produce any nuclide lighter than the target nucleus.

Charged particles with energies lower than the Coulomb barrier (a few MeV) do not react effectively with nuclei. As soon as the energy exceeds the Coulomb barrier, compound nuclei may be formed, which de-excite by the emission of photons, nucleons, or light nuclei (e.g., in the case of protons, reactions of type (p, n), (p, d), (p, α), etc. can occur). Similarly to neutrons, high-energy charged particles interact by spallation reactions.

Electromagnetic particles may also cause activation through photonuclear interactions, although with a much lower cross-section than for hadronic reactions (at high energy, lower by the fine structure constant). Thus, activation by electrons and photons is typically not a concern at hadron accelerators, whereas it might be important at electron accelerators. The threshold energies for photonuclear reactions are a few MeV, depending on the target material. Just above threshold, so-called giant dipole resonance reactions dominate, in which the nucleus de-excites by the emission of neutrons, protons, and light nuclei.

### 7.1 Fundamental principles

Radioactive decay is a random process characterized by a decay constant $\lambda$. If a total number $N_{\text{tot}}(t)$ atoms of a radionuclide are present at time $t$, the total activity $A_{\text{tot}}(t)$ is determined by

$$A_{\text{tot}}(t) = \frac{dN_{\text{tot}}(t)}{dt} = \lambda N_{\text{tot}}(t), \qquad (9)$$

for which the solution at $t = T$ is

$$A_{\text{tot}}(T) = A_{\text{tot}}(0) e^{-\lambda T}. \qquad (10)$$

Often, the time required to decay to half of the original activity, the half-life $t_{1/2}$, is given; this is related to the decay constant by

$$t_{1/2} = \frac{\ln 2}{\lambda}. \qquad (11)$$

If we assume steady irradiation of a material with a spatially uniform fluence rate $\Phi$ (cm$^{-2} \cdot$s$^{-1}$), the density of atoms $n(t)$ of the radionuclide of interest per unit volume at time $t$ (cm$^{-3}$) during the irradiation is governed by

$$\frac{dn(t)}{dt} = -\lambda n(t) + N\sigma\Phi, \qquad (12)$$

where $\sigma$ is the production cross-section (cm$^2$) and $N$ is the density of target atoms (cm$^{-3}$). This equation has the solution

$$n(t) = \frac{N\sigma\Phi}{\lambda}(1 - e^{-\lambda t}), \qquad (13)$$

where the specific activity during irradiation is given by $A(t) = \lambda n(t)$. For $t \gg t_{1/2}$, Eq. (13) yields $A(t) = A_{\text{sat}} = N\sigma\Phi$, i.e. the saturation activity equals the production rate.

The activity after an irradiation period $t$ and a cool-down time $t_{cool}$ can be written as

$$A_{tot}(T) = A_{sat}(1 - e^{-t/\tau})e^{-t_{cool}/\tau}, \tag{14}$$

where $\tau = 1/\lambda$.

## 7.2 Radionuclides in solid materials

The most important medium- and long-lived radionuclides produced in typical accelerator materials are given in Table 4. As can be seen, the heavier the elements in the material are, the greater the number of radionuclides that can be created. Thus, light materials should be preferred if possible in the construction of accelerator components. For example, aluminium supports have better radiological characteristics than steel supports owing to the significantly lower number of nuclides produced.

Reactions with trace elements in materials give rise to additional nuclides which might also be important, especially if they are long-lived. A typical example is $^{60}$Co, produced by thermal-neutron capture reactions with traces of cobalt in aluminium or iron components. This nuclide can dominate the activity in a component many years after irradiation, when most other nuclides have already decayed.

The activation properties of the materials used in accelerator construction must be considered during the design process as they may have a direct impact on later handling (maintenance and repair) and waste disposal. Gamma-emitting nuclides dominate the residual dose rates at longer decay times (more than one day), whereas at short decay times $\beta^+$ emitters are also important (as a result of photons produced by $\beta^+$ annihilation). Owing to their short range, $\beta^-$ emitters are usually relevant only to doses to the skin and eyes and doses due to inhalation or ingestion.

Figures 4 and 5 show the contributions of gamma and $\beta^+$ emitters, respectively, to the total dose rate close to an activated copper sample [12]. Typically, the dose rates at a given decay time are determined mainly by radionuclides with half-lives of the order of the decay time. Extended irradiation periods might be an exception to this general rule, as in this case the activity of long-lived nuclides can build up sufficiently that it dominates over that of short-lived nuclides even at short cooling times.

Activation in concrete is dominated by $^{24}$Na (at short decay times) and $^{22}$Na (at long decay times). Both of these nuclides can be produced either by low-energy neutron reactions with the sodium component in the concrete or by spallation reactions with silicon and calcium. At long decay times, the nuclides of radiological interest in activated concrete can also include $^{60}$Co, $^{152}$Eu, $^{154}$Eu, and $^{134}$Cs, all of which are produced by (n, γ) reactions with traces of natural cobalt, europium, and caesium. Thus, such trace elements might be important even if their content in the concrete is only a few parts per million or less by weight.

Explicit simulation of radionuclide production with general-purpose Monte Carlo codes has become the method most commonly applied to calculate induced radioactivity and its radiological consequences. Nevertheless, other more approximate approaches, such as the use of '$\omega$-factors' [13], can still be useful for fast order-of-magnitude estimates. These $\omega$-factors give the dose rate per unit star density (the density of inelastic reactions above a certain energy threshold, e.g. 50 MeV) in contact with an extended, uniformly activated object after 30 days of irradiation and one day of decay. The $\omega$-factor for steel or iron is approximately $3 \times 10^{-12}$ Sv cm$^3$/star. This does not include possible contributions from thermal-neutron activation.

**Table 4:** Nuclides of radiological importance in the elements of typical accelerator materials. The last column indicates the half-life.

| Element or material | Nuclide | $t_{1/2}$ |
|---|---|---|
| Carbon | $^3$H | 12.3 y |
| | $^7$Be | 53.29 d |
| | $^{11}$C | 20.38 min |
| Aluminium | All of the above plus | |
| | $^{22}$Na | 2.6 y |
| | $^{24}$Na | 15.0 h |
| Iron | $^{m44}$Sc | 2.44 d |
| | $^{46}$Sc | 83.8 d |
| | $^{48}$Sc | 1.81 d |
| | $^{48}$V | 16.0 d |
| | $^{51}$Cr | 27.7 d |
| | $^{52}$Mn | 5.6 d |
| | $^{54}$Mn | 312.1 d |
| | $^{55}$Fe | 2.73 y |
| | $^{59}$Fe | 44.5 d |
| | $^{55}$Co | 17.54 h |
| | $^{56}$Co | 77.3 d |
| | $^{57}$Co | 271.8 d |
| | $^{58}$Co | 70.82 d |
| Stainless steel | All of the above plus | |
| | $^{60}$Co | 5.27 y |
| | $^{57}$Ni | 35.6 h |
| Copper | All of the above plus | |
| | $^{63}$Ni | 100 y |
| | $^{61}$Cu | 3.4 h |
| | $^{64}$Cu | 12.7 h |
| | $^{65}$Zn | 244.3 d |

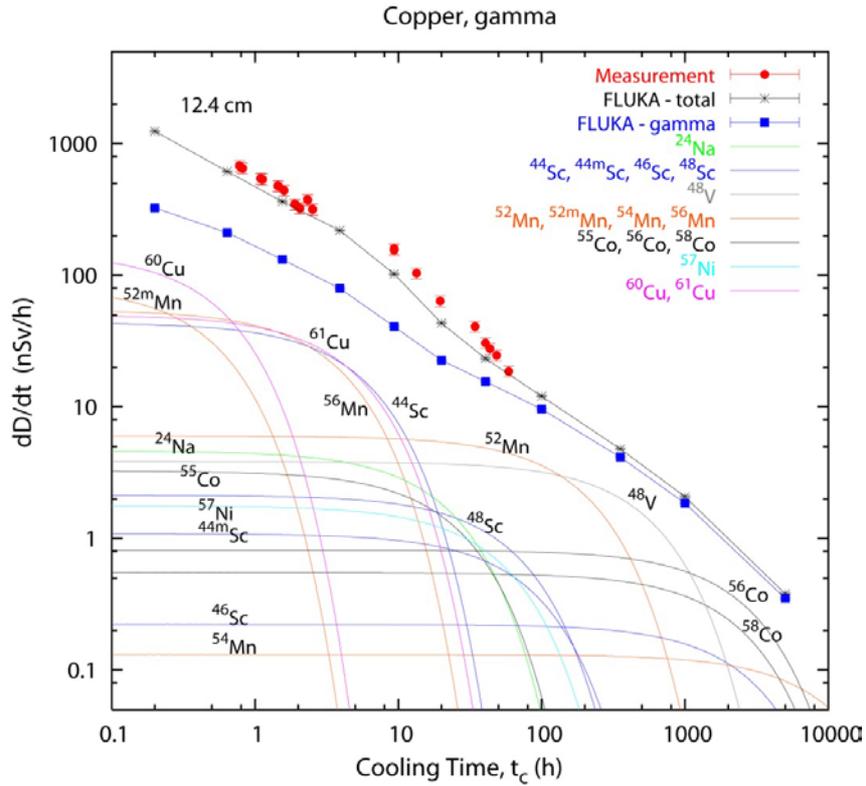

**Fig. 4:** Contribution of individual gamma-emitting nuclides to the total dose rate at 12.4 cm from an activated copper sample [12]

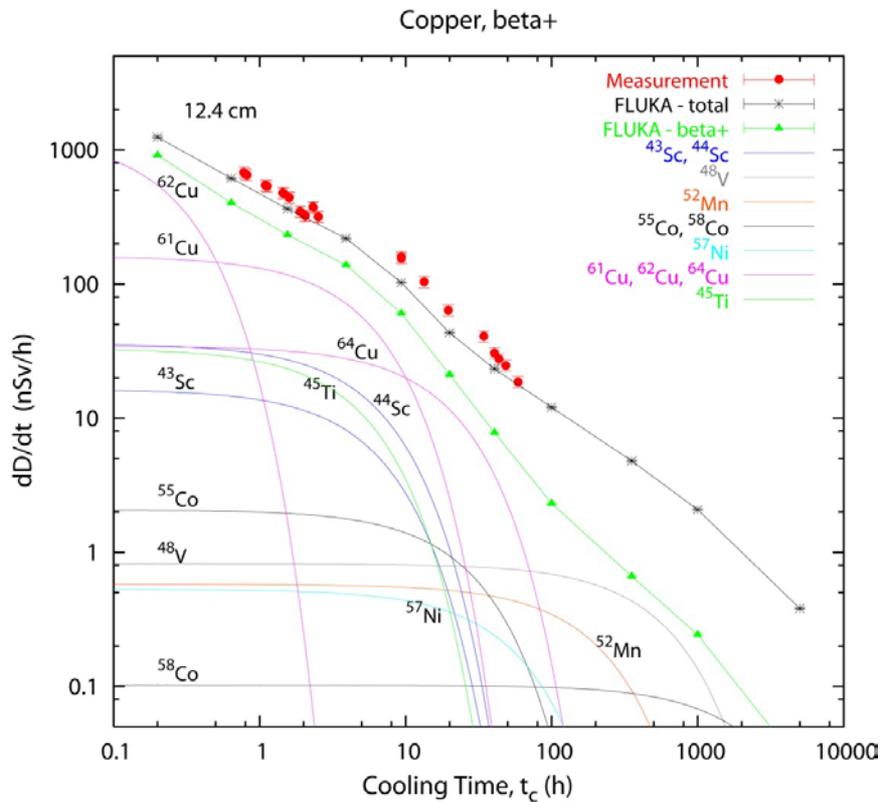

**Fig. 5:** Contribution of individual positron-emitting nuclides to the total dose rate at 12.4 cm from an activated copper sample [12]

## 7.3 Radionuclides in liquids

At accelerators, liquids are used mainly for cooling purposes (e.g. demineralized water and liquid helium), but liquid targets also exist (e.g. mercury).

Spallation reactions of secondary particle showers with oxygen in demineralized water can create tritium ($t_{1/2}$ = 12.3 y), $^7$Be ($t_{1/2}$ = 53.29 d), and a number of short-lived $\beta^+$-emitters ($^{11}$C, $^{13}$N, and $^{15}$O). The production of tritium by thermal-neutron capture in natural hydrogen can be neglected in most application owing to the low abundance of deuterons and the small cross-section. Sometimes cooling-water circuits also contain nuclides from corrosion products (e.g. cobalt nuclides); however, a large fraction of these is collected, together with $^7$Be, in the resin of ion exchanger cartridges. In natural water, radionuclides can also be produced in reactions with trace elements (i.e. minerals).

During accelerator design, the activation of cooling liquids is most conveniently assessed by folding fluence spectra with energy-dependent nuclide production cross-sections. Direct calculation is also possible using Monte Carlo codes for nuclides produced from oxygen, but this direct method would fail for nuclides produced from trace elements owing to a lack of statistical significance.

Activated cooling liquids pose contamination hazards during interventions in accelerator components and may also cause external irradiation close to pipes and cartridges. Although the decay of tritium proceeds only via the emission of a low-energy electron, its concentration in water, especially if released off-site, has become a critical parameter as it may attract the attention of the public.

## 7.4 Radionuclides in air

Airborne radionuclides are produced mainly by the interaction of beam particles or associated showers of secondary particles with air molecules. Other sources include activated dust and outgassing of nuclides from activated accelerator components. The latter two sources, however, are typically of lower importance and can only be assessed by measurement.

Table 5 gives the nuclides of highest radiological importance. At hadron and ion accelerators, most of them are created by spallation reactions with air molecules. Only $^{41}$Ar results from thermal-neutron capture reactions with argon ($\sigma_{th}$ = 660 mb). At electron accelerators, photonuclear interactions of type ($\gamma$, n) contribute to the production of $^{13}$N and $^{15}$O. Although the radiological impact of $^3$H in air is small, it easily becomes attached to humidity and can reach waste water circuits, especially via condensation in air conditioning units.

**Table 5:** Airborne nuclides of radiological importance (the second column indicates the half-life)

| Nuclide | $t_{1/2}$ |
|---|---|
| $^3$He | 12.3 y |
| $^7$Be | 53.29 d |
| $^{11}$C | 20.38 min |
| $^{13}$N | 9.96 min |
| $^{15}$O | 2.03 min |
| $^{41}$Ar | 1.83 h |

Apart from the list in Table 5, specific situations and exposure pathways may require the consideration of further nuclides, such as $^{32}$P ($t_{1/2}$ = 14.26 d), which is produced by spallation reactions with argon. This nuclide can reach milk consumed by infants through ground deposition on grazing land and thus dominate the committed dose due to ingestion.

The low density of air usually renders a direct calculation of the activation of air by Monte Carlo models highly inefficient. Instead, particle fluence spectra are multiplied by energy-dependent nuclide production cross-sections, which are obtained from Monte Carlo models, experimental data, or both (the latter are called evaluated cross-sections). This yields nuclide production rates per unit volume or, after application of Eq. (13), the specific activity.

The results of air activation studies play a crucial role in the design of the ventilation system of an accelerator. Closed circuits that are flushed with fresh air prior to access but otherwise remain closed have the advantage of reducing the total annual release of short-lived nuclides. However, the concentration of long-lived nuclides may build up and lead to undue exposure if the nuclides are released at once over a period of time too short for there to be any benefit from changing wind conditions. In addition, tritium can build up, attach to water, and accumulate, for example in sumps. On the other hand, constant venting with fresh air causes an increased annual release of short-lived nuclides, although there is a benefit from natural dilution of long-lived nuclides. Apart from the environmental aspects, ventilation systems have safety functions in ensuring the containment of radioactive gases and should follow international standards [14].

Adjustments for the presence of ventilation can be made by introducing an effective decay constant $\lambda'$ that includes the physical decay constant along with a ventilation term:

$$\lambda' = \lambda + \frac{D}{V}, \tag{15}$$

where $D$ is the ventilation rate (volume of air exchanged per unit time) and $V$ is the enclosure volume. Thus, with ventilation, the saturation activity $A'_{\text{sat}}$ becomes

$$A'_{\text{sat}} = \frac{\lambda A_{\text{sat}}}{\lambda + D/V}. \tag{16}$$

## 8 Radiation fields around high-energy accelerators

### 8.1 Prompt stray radiation fields

Stray radiation fields are created at high-energy particle accelerators by the intentional interaction of the accelerated beam with targets, beam dumps, and collimators and by unintentional beam losses on structural components of the machine.

At electron accelerators, the most important secondary radiation is bremsstrahlung photons and high-energy electrons produced in electromagnetic cascades. An electromagnetic cascade is initiated when either a high-energy electron or a high-energy photon enters a material. At high energy, photons interact with matter mainly via pair production, whereas electrons and positrons lose their energy in a medium primarily by emitting bremsstrahlung photons. These two processes continue alternately, leading first to an exponential increase in the number of particles present in the cascade, which then starts to decline when removal processes (the photoelectric effect, ranging-out of electrons, and Coulomb and Compton scattering) dominate over the processes that generate new particles. Finally, low-energy electrons lose their residual energy by ionization and excitation processes.

At high-energy electron accelerators, neutrons are also present, released by photon-induced reactions rather than by electrons directly. High-energy neutrons are often the dominant secondary radiation outside a thick shield, which usually absorbs most of the bremsstrahlung photons.

At proton accelerators, interaction of the beam with materials generates a hadron cascade containing neutrons, charged hadrons, muons, photons, and electrons, with energy spectra extending over a wide range. The number of secondary particles produced per primary proton (the multiplicity) increases as the proton energy increases. The average energy of these secondary particles also increases with the energy of the primary proton, making them capable of producing further inelastic interactions. The dominant radiation at workplaces outside accelerator shielding is the neutron field, with minor contributions from other particles. The neutron spectrum at the source, for example a beam loss point, is modified by transport through the shield, so that the energy distribution of neutrons at a workplace may be significantly different from the source spectrum. The shape of the spectrum also depends on the thickness of the shielding: the various components of the spectrum are attenuated differently, and only after a certain depth in the shield does the neutron spectrum reach equilibrium. This can be seen in Figs. 6 and 7, which show the neutron energy distributions in the transverse direction generated by 250 MeV protons impinging on an iron target thicker than the proton range. These figures show the energy distribution of the source neutrons and that behind a thin (20 cm to 1 m) and a thick (1–5 m) concrete shield. The distributions have been normalized to unit area in order to show better the change in the shape of the spectrum with increasing shield thickness.

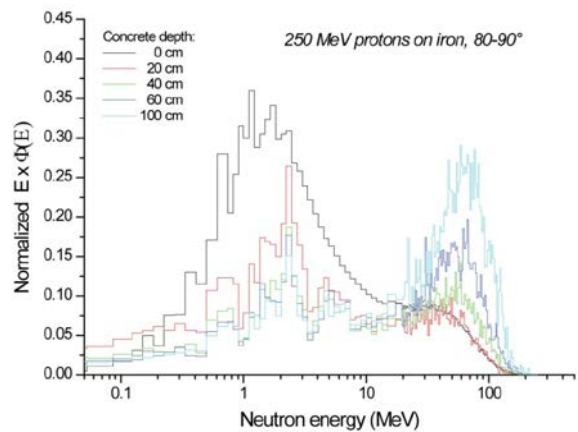

**Fig. 6:** Neutron energy distributions $E\Phi(E)$ in the transverse direction generated by 250 MeV protons impinging on an iron target thicker than the proton range. The distributions are for source neutrons and behind concrete shields of thicknesses ranging from 20 cm to 1 m. The distributions have been normalized to unit area in order to show better the change in the shape of the spectrum with increasing shield thickness.

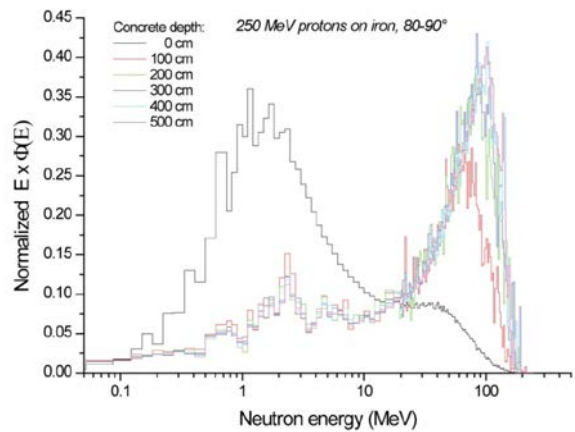

**Fig. 7:** Neutron energy distributions $E\Phi(E)$ in the transverse direction generated by 250 MeV protons impinging on an iron target thicker than the proton range. The distributions are for source neutrons and behind concrete shields of thicknesses ranging from 1 m to 5 m. The distributions have been normalized to unit area in order to show better the change in the shape of the spectrum with increasing shield thickness.

Figure 8 shows typical neutron energy distributions outside two types of shield at a multi-GeV proton accelerator [15]. The difference between the shapes of the two spectra outside the concrete shields is because in one case the neutrons emerging from the shield are scattered further by an additional surrounding concrete structure which softens the spectrum, a situation commonly found at accelerators.

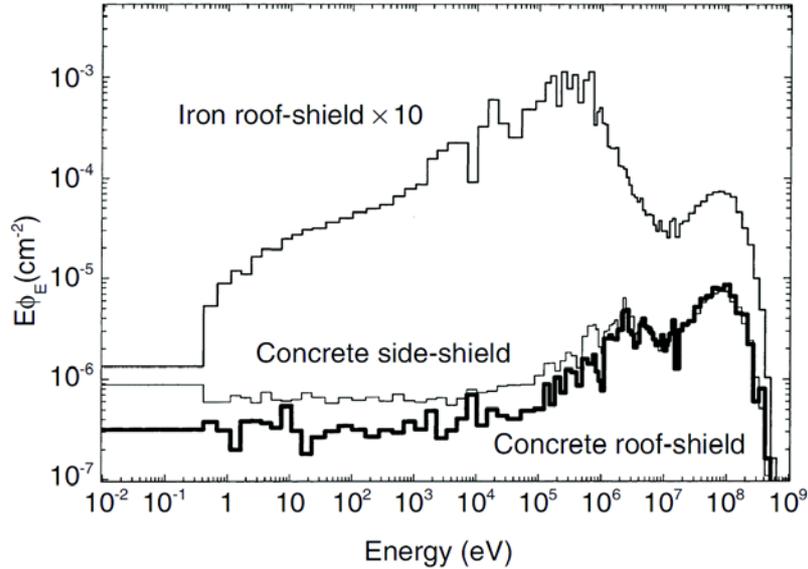

**Fig. 8:** Neutron spectral fluences $E\Phi(E)$ outside a concrete roof shield (80 cm thickness of concrete), an iron roof shield (40 cm thickness of iron), and an 80 cm thick concrete side shield (80 cm thickness of concrete, but the neutrons are scattered further by surrounding concrete) at the CERF facility at CERN (neutrons per primary beam particle incident on a copper target) [15]

As an example of the contribution of particles other than neutrons to $H^*(10)$, Figs. 9 and 10 plot the ratio of the values of $H^*(10)$ due to protons, photons, and electrons at various depths in a concrete shield to the total, in the forward and transverse directions, for 250 MeV protons impinging on a thick iron target. One sees that in the forward direction, protons contribute more than photons, while in the transverse direction, the opposite is the case.

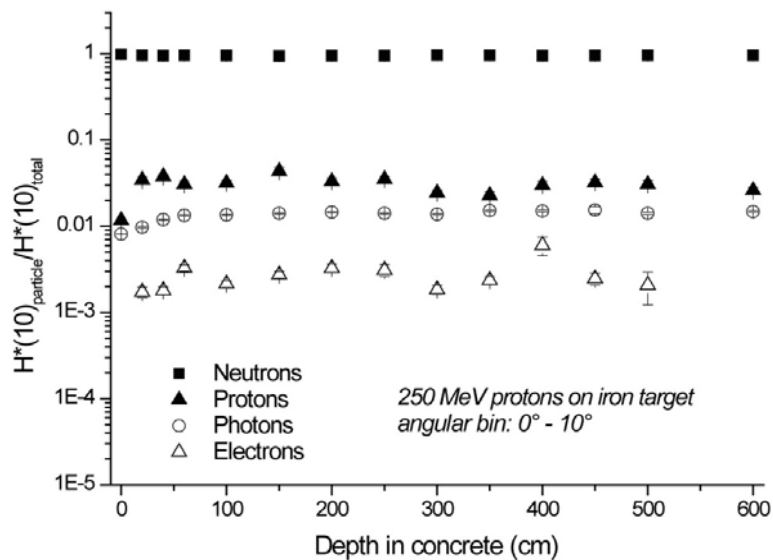

**Fig. 9:** Ratio of $H^*(10)$ due to secondary particles at various depth in a concrete shield to the total, in the forward direction, for 250 MeV protons impinging on an iron target thicker than the proton range.

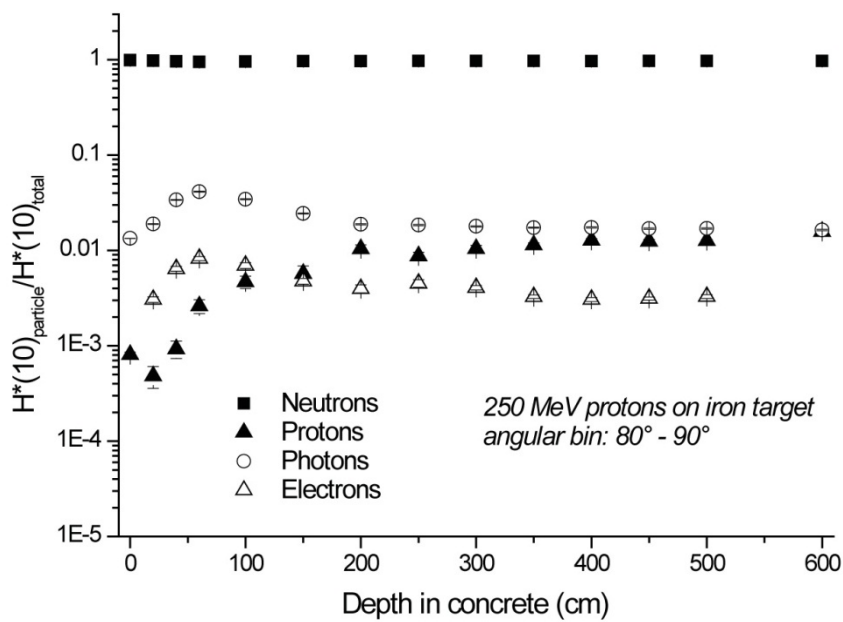

**Fig. 10:** Ratio of $H^*(10)$ due to secondary particles at various depth in a concrete shield to the total, in the transverse direction, for 250 MeV protons impinging on an iron target thicker than the proton range

Above about 10 GeV, muon-shielding requirements dominate in the forward direction for high-intensity proton beams, meaning that a residual muon beam is often present behind a shield thick enough to attenuate the hadron component of the field [16]. Muons arise from the decay of pions and kaons, either in the particle beam or in cascades induced by high-energy hadrons [17]. They can also be produced in high-energy hadron–nucleus interactions. The decay lengths for pions and kaons are 55.9 m and 7.51 m, respectively, times the momentum (in GeV/*c*) of the parent particle. Muons are weakly interacting particles and can only be stopped by 'ranging them out'. Muons lose energy mainly by ionization, as their cross-section for nuclear interaction is very low.

Muons from pion decay have a momentum spectrum that extends from 57% of the momentum of the parent pion to the pion momentum itself. Secondary pion beams generally have dumps containing a longitudinal depth of 1–2 m of Fe, and thus decay muons will penetrate these dumps for pion beams with a momentum larger than a few GeV/*c*.

To give an example [17], a beam of $10^7$ pions per pulse with a momentum of 20 GeV/*c* travelling over a distance of 50 m will generate about $5 \times 10^5$ muons per pulse (5% of the parent beam). For a pulse repetition period of 2 s (a typical order of magnitude for a high-energy synchrotron), taking an approximate fluence-to-dose-equivalent conversion factor equal to 40 fSv·m$^2$ [18] and assuming that the muon beam is averaged over a typical area for a human torso of 700 cm$^2$, this fluence translates into a non-negligible dose equivalent rate of 500 µSv/h. Thus, under some circumstances (e.g. if the area downstream of the beam line is not interlocked for access), a muon component can be present in a mixed workplace field and contribute substantially to personnel exposure.

Radiation protection quantities such as the dose rate at workplaces and shielding thickness are generally not simple functions of energy. The parameters which most directly affect radiological safety are the particle type, the particle energy, the average beam power, and the number of lost particles per unit time at a given energy.

Some accelerators operate in pulsed mode, which means that the beam is present in the machine (or lost somewhere) during only a fraction of the time. With a synchrotron, the relevant parameters are the repetition rate (the number of cycles per unit time) and the flat-top duration (the time during which the beam is extracted from the accelerator to be transported somewhere else), whereas a cyclotron produces a virtually continuous beam. With a linear accelerator, an important parameter is the duty factor (DF), which is the fraction of the operating time during which the linac is actually producing radiation:

$$\text{DF} = p * T_p, \tag{17}$$

where $p$ is the pulse repetition rate (in Hz) and $T_p$ is the pulse length (in seconds).

At a given energy $E$, the dose rate generated by the interaction of the beam with a material is directly proportional to the average beam power $P$ (i.e. to the number of 'lost' particles).

### 8.2  Stray radiation from residual radioactivity

Residual radioactivity is mainly a problem at proton accelerators, as dose rates at electron machines from induced radioactivity in accelerator structures are typically two orders of magnitude lower. At most accelerator facilities, the largest contribution to personnel dose actually arises from maintenance work near dumps, targets, septa, and collimators and generally near any object hit directly by the primary beam or located close to a beam loss point, rather than from exposure during machine operation. External (and sometimes internal) exposure to radiation from induced radioactivity can also occur in connection with the handling, transport, machining, welding, chemical treatment, and storage of irradiated items. A place where personnel can be exposed to such types of radiation can also be a workplace.

In spite of the fact that this radiation source is actually responsible for most of the individual and collective doses at accelerator laboratories, the associated radiation field is much simpler than that of the prompt radiation generated during accelerator operation, and the personnel exposure is due only to beta- and gamma-emitting radionuclides (and whole-body exposure is due essentially only to gamma emitters). The most common radionuclides with sufficiently long half-lives found in accelerator components are $^7$Be, $^{22}$Na, $^{54}$Mn, $^{65}$Zn, and the cobalt isotopes $^{56}$Co, $^{57}$Co, $^{59}$Co, and $^{60}$Co; in activated shielding structures, $^{133}$Ba, $^{134}$Cs, and $^{137}$Cs are found; and in earth used as a shielding material, $^{152}$Eu and $^{154}$Eu are found.

The monitoring of such workplaces thus requires only beta/gamma monitors, such as ion chambers.

## 9  Instrumentation for area monitoring

CERN has a legal obligation to protect the public and persons working on its site from any unjustified exposure to ionizing radiation. For this purpose, CERN's Occupational Health & Safety and Environmental Protection (HSE) Unit monitors ambient dose equivalent rates inside and outside CERN's perimeter and releases of radioactivity in air and water. The results of the measurements allow the preventive assessment of radiological risks and the minimization of individual and collective doses. CERN's HSE Unit currently operates two radiation monitoring systems:

– ARCON (ARea CONtroller), which was developed at CERN for LEP and has been in use since 1988;
– RAMSES (RAdiation Monitoring System for the Environment and Safety), which was designed for the LHC based on current industry standards and has been in use since 2007.

About 800 monitors are employed in ARCON and RAMSES, about 400 for each system. Both installations comprise data acquisition, data storage, and the triggering of radiation alarms and beam interlocks. The most recent CERN facilities (the LHC, CNGS, and CTF3) are equipped with RAMSES, whereas the entire LHC injector chain, the remaining facilities (e.g. ISOLDE, n-TOF, and AD), and all experimental areas are still equipped with ARCON. In the long run, it is envisaged that ARCON will be replaced by the more recent RAMSES technology.

## 9.1 Radiation monitors

Both ARCON and RAMSES use the same or at least very similar types of radiation detectors. Environmental radiation protection monitors record stray radiation and the releases of radioactivity into air and water. Recording of other measured values such as wind speed, wind direction, and flow rates is required to obtain relevant input parameters for calculating doses to members of the public. An environmental stray-radiation monitoring station consists of one high-pressure ionization chamber filled with argon (from Centronics) for photons and penetrating charged particles such as muons, one REM counter (from Berthold) for neutrons, and a locally installed unit for data acquisition, alarm generation, and data transfer. The radiation protection part of a CERN water monitoring station consists of an NaI detector for in-situ measurements of gamma-emitting radionuclides and a device to collect water samples for laboratory analyses such as measurements of tritium and for cross-checks of the on-line results. The ventilation monitoring system is based on silicon surface detectors to measure the total activity of beta emitters released. In addition, removable filters are installed to allow laboratory analysis of radionuclides attached to aerosols using gamma spectroscopy. The active parts of the air and water monitoring stations (the Si and NaI detectors) are equipped with alarm functions.

The Radiation Protection Group uses three different types of monitors to measure ambient dose equivalent rates at CERN and in the close neighbourhood of CERN's facilities. The radiation monitors employed to protect workers against prompt ionizing radiation [19] during beam operation are special REM counters (from WENDI/Thermo) and hydrogen-filled, high-pressure ionization chambers (from Centronics). Both are optimized to measure high-energy neutrons with energies up to the GeV range; the hydrogen chamber responds to all particles contributing to the high-energy mixed radiation fields [20, 21].

The ambient dose equivalent rates which can be monitored inside the machine tunnel and the experimental caverns after the beam has been stopped are due to radiation emitted by the decay of radionuclides induced during operation of the beam. The energies of the emitted photons do not exceed 2.7 MeV (emitted by $^{24}$Na) [19]. The induced radioactivity is measured with air-filled plastic ionization chambers (from PMI) in order to assess risks during maintenance and repair work [22]. The radiation monitoring system is completed by hand and foot monitors at the exits from the accelerator and experimental areas and by gate monitors at the exits of the CERN sites (Site Gate Monitors, SGMs). The RAMSES system provides an option to connect the SGMs to the access system; that is, when there is an alarm, the barriers can remain closed.

Outside the shield of an accelerator facility, the ambient dose equivalent rates during operation range from a few hundreds of microsieverts per year to a few millisieverts per year. To measure such rates, one needs detectors that are of high sensitivity or capable of integrating over long periods.

## 9.2 Dosimetry at CERN

Exposure to ionizing radiation (gamma, beta, and particle radiation) accompanies all work at a particle accelerator and in the associated experimental facilities. Legal dose limits assure the safety of personnel working under these conditions. The dose received by individuals working with ionizing radiation at CERN is monitored with personal dosimeters. Every person working at CERN in Radiation Areas or with sources of ionizing radiation must wear a CERN dosimeter. The CERN dosimeter registers the personal dose from sources of ionizing radiation around particle accelerators. It combines an active detector for gamma and beta radiation based on the Direct-Ion Storage (DIS) technology and a passive detector for quantifying neutron doses.

The gamma/beta dose registered by a CERN dosimeter can be read out as frequently as deemed necessary, but it must be read at least once per month on one of the approximately 50 reader stations which are installed CERN-wide The monitoring period for the neutron dosimeter is, in principle, one year. It must then be returned to the supplier for evaluation.

For work in Controlled Radiation Areas, where the radiological risk and the dose rate are above 50 µSv/h, the additional use of an operational dosimeter is required. CERN provides all staff who may work in Limited Stay Radiation Areas or High Radiation Areas with a system for active dosimetry with an alarm, in the form of an dosimeter, model DMC-2000 from MPG instruments.

## 10  ALARA at CERN

CERN introduced a formalized approach to ALARA [23–25] at the end of 2006, as a result of collaboration between the former Accelerator and Beams department and the Radiation Protection Group. This approach was applied first to the SPS and LHC complex, and since 2009 has been applied to all CERN facilities. The goal was to optimize work coordination, work procedures, handling tools, and even the design of entire facilities. Consequently, all work in Radiation Areas has to be optimized. In particular, all work in Controlled Radiation Areas must be planned and optimized, including an estimate of the collective and individual effective doses to the workers participating in the completion of a task.

Five different criteria were established in 2006 and are used for the determination of the so-called ALARA level of an intervention. These five criteria are shown in Table 6. Depending on the level of the intervention, different means of optimization have to be applied. For example, level 3 interventions need formal approval from the ALARA Committee, which is chaired by the Director of Accelerators.

**Table 6:** ALARA criteria at CERN

Criteria: Ambient dose equivalent

| Level I | Level II | Level III |
|---|---|---|
| | 50 µSv/h | 2 mSv/h |

Criteria: Individual dose

| Level I | Level II | Level III |
|---|---|---|
| | 100 µSv | 1 mSv |

Criteria: Collective dose

| Level I | Level II | Level III |
|---|---|---|
| | 500 µSv | 10 mSv |

Criteria: Airborne activity in CA values according to [26]

| Level I | Level II | Level III |
|---|---|---|
| | 5 CA | 200 CA |

Criteria: Surface contamination in CS values according to [26]

| Level I | Level II | Level III |
|---|---|---|
| | 10 CS | 100 CS |

**Acknowledgements**

Figures 6, 7, 9, and 10 are the results of Monte Carlo simulations performed at CERN by Alessio Mereghetti for his diploma thesis at the Polytechnic of Milan. The authors wish to thank Alessio for kindly providing the figures for this paper.